\documentstyle[12pt]{article}
\def\hybrid{\topmargin -20pt    \oddsidemargin 0pt
        \headheight 0pt \headsep 0pt
        \textwidth 6.25in       
        \textheight 9.5in       
        \marginparwidth .875in
        \parskip 5pt plus 1pt   \jot = 1.5ex}

\hybrid
\newskip\humongous \humongous=0pt plus 1000pt minus 1000pt
\def\caja{\mathsurround=0pt}
\def\eqalign#1{\,\vcenter{\openup1\jot \caja
        \ialign{\strut \hfil$\displaystyle{##}$&$
        \displaystyle{{}##}$\hfil\crcr#1\crcr}}\,}
\newif\ifdtup

\relax

\def\be{\begin{equation}}
\def\ee{\end{equation}}
\def\ba{\begin{eqnarray}}
\def\ea{\end{eqnarray}}

\begin{document}
\renewcommand{\theequation}{\thesection.\arabic{equation}}
\newcommand{\beq}{\begin{equation}}
\newcommand{\eeq}[1]{\label{#1}\end{equation}}
\newcommand{\ber}{\begin{eqnarray}}
\newcommand{\eer}[1]{\label{#1}\end{eqnarray}}
\begin{titlepage}
\begin{center}

\hfill CPTh-A295.02.94\\
\hfill Crete-94-10\\
\hfill hep-th/9403034\\
\hfill \\

\vskip .5in

{\large \bf  RIBBONS AROUND MEXICAN HATS }
\vskip .5in

{\bf C. Bachas} \footnotemark \\

\footnotetext{e-mail address: bachas@orphee.polytechnique.fr}

\vskip .1in

{\em Centre de Physique Th\'eorique\\
Ecole Polytechnique \\
 91128 Palaiseau, FRANCE}

\vskip .15in

       and

\vskip .15in

{\bf T.N. Tomaras} \footnote{e-mail address:
tomaras@iesl.forth.gr }\\
\vskip
 .1in

{\em  Physics Dept., University of Crete\\
and Research Center of Crete\\
 714 09 Heraklion, GREECE
  }\\

\vskip .1in

\end{center}

\vskip .4in

\begin{center} {\bf ABSTRACT }
\end{center}
\begin{quotation}\noindent
 We analyze quasi-topological solitons
  winding around a mexican-hat
 potential in two space-time
dimensions.  They are prototypes for a large
 number of physical excitations,
including  Skyrmions  of the Higgs sector of the standard
electroweak  model,
magnetic bubbles in thin ferromagnetic films,
 and strings in certain non-trivial
backgrounds.
 We present explicit solutions, derive the conditions
for classical stability, and show that
contrary to the naive expectation
these can be satisfied in
the  weak-coupling limit.
In this limit we can calculate the soliton
 properties reliably, and
 estimate their lifetime semiclassically.
 We explain why
 gauge interactions destabilize these solitons, unless
the scalar sector is extended.

\end{quotation}
\vskip1.0cm
February 1994\\
\end{titlepage}
\vfill
\eject
\def\baselinestretch{1.2}
\baselineskip 16 pt
\noindent
\section{Introduction}
\setcounter{equation}{0}

\hskip 0.6cm  In many different physical
 contexts there arise solitonic excitations
that are characterized by the non-trivial
wrapping of a $d$-dimensional
target sphere by $d$-dimensional space.
Examples include Skyrmions
 \cite{Skyrme}
of the chiral pion Lagrangian or of the
 Higgs sector in the electroweak
model, magnetic bubbles in thin ferromagnetic films
\cite{bubbles}, and
winding states in closed-string compactifications
\cite{CS}.
Such solitons have up to now been discussed only in the context of
 non-linear $\sigma$-models,
with fields living on the $d$-sphere, or more generally on a
target manifold
${\cal M}$ whose homotopy group $\pi_d({\cal M})$ is non-trivial.
In this paper we study the fate of these solitons, when
the radial degree of freedom
of the fields is unfrozen.
In the above
examples this would correspond to allowing
the physical Higgs field, the
$\sigma$-field in the Gell-Mann-Levy model, or the magnitude of the
magnetization to fluctuate. The existence and stability of solitons
becomes in this case a dynamical issue, and does not follow
from simple topological considerations.

In contrast to other non-topological
 solitons  \cite{Q} such as $Q$-balls,
our solitons are static and owe their (meta)stability to the
dynamical exclusion of some region of target space.
They are easiest to visualize in one spatial
dimension, as ribbons tied around a mexican hat.
In this paper we will mostly restrict ourselves to this simplest
prototype model, in which the quantitative analysis can be carried
the furthest. We will show that winding solitons
  are classically stable
provided the  mass of the radial field,
measured in units of their size, exceeds some critical value.
Contrary to the naive expectation,
 this condition can be satisfied even in the weak-coupling
limit, in which furthermore the solitons' properties
 can be   reliably calculated
  within a semiclassical approximation.
Winding solitons decay eventually
by slipping off the tip of the hat, and we will identify one instanton
that contributes to this tunneling process.
 A novel instability   arises
in the presence of gauge interactions, but can as we will
explain be completely suppressed in models with an
extended scalar sector.

The consequences of our results for the physics of Skyrmions
and of magnetic vortices will be discussed in a
  separate publication
\cite{BT}. Our classical analysis
will in fact carry through with minor modifications only.
Only in one spatial dimension can we, however, at present affirm the
existence of such winding solitons at
 a fully controllable quantum level.

\def\baselinestretch{1.2}
\baselineskip 16 pt
\noindent
\section{Winding Solitons }

\hskip 0.6cm  Consider a complex scalar field
 in two space-time dimensions with
action
 $$ S = \int dt\ dx \Bigl[ {1\over 2}
 \partial_\mu \Phi^* \partial^\mu \Phi -
{\lambda\over 4}
(\Phi^* \Phi - v^2)^2 \Bigr] \ .\eqno(1)$$
Its perturbative spectrum consists of a
massless Goldstone boson of the
broken $U(1)$ symmetry, and a radial degree of freedom with mass
$m = \sqrt{2\lambda} v$.
The solitons we will consider are characterized by the non-trivial
winding of the complex field around zero.
They are the analog of
three-dimensional Skyrmions, whose distinguishing feature is the
wrapping of the target three-sphere  by
the pion or the Higgs-doublet field
configuration.
A major difference between one and
three space dimensions should however be kept in mind.
  In three dimensions the finite-energy requirement
  compactifies space automatically
since the fields must go to a constant at infinity.
 Classical winding
configurations on the other hand
  have  a tendency to shrink to zero size,
 and must be stabilized at some scale $\rho$
by higher-derivative
terms in the (effective) action \cite{Skyrme}.
 In one space dimension
 there is no shrinking instability.
On the other hand
 for theories like model (1)  with a continuously degenerate
ground state
 space is not compactified automatically,
 since the field need not tend
to the same value  at $\pm\infty$.
In order to get non-trivial topology we must either
lift the vacuum degeneracy by introducing
 a small mass  $\delta V = \mu^2
Re(\Phi^2)$  in the angular direction,
or else compactify space by hand by imposing periodic
boundary conditions: $x\equiv x+2\pi L$. Both $L$ and $\mu^{-1}$
are infrared cutoffs, which prevent the soliton from stretching to
infinite size, and play a role analogous to the
effective ultraviolet cutoff of the Skyrme model. We prefer for
technical convenience to use $L$ rather
than $\mu^{-1}$ in the sequel,
but this choice should not affect our conclusions in any
important way.

Let us then concentrate on model (1) defined on compactified space.
In the $\lambda\rightarrow\infty$ limit
the  complex field   is forced to lie on a circle and there exists a
conserved topological current
$$ J^\mu = {1\over 2\pi i}
 \epsilon^{\mu\nu} \partial_\nu \log\Phi \ , \eqno(2) $$
whose corresponding charge $N = \int dx\  J^0 $ is the
winding number.
When the magnitude of the field
is allowed to fluctuate the conservation law
gets modified to
$$ \partial_\mu J^\mu = {\cal J}
\ \delta^{(2)} \Bigl( \Phi(x,t) \Bigr) \ , \eqno(3a) $$
where
$$  {\cal J}
= {1\over 2i} \epsilon_{\mu\nu}
 \partial_\mu\Phi^* \partial_\nu \Phi \  \eqno(3b)
$$
is the Jacobian of the mapping of
 space-time onto the complex $\Phi$-plane. These
topological equations express the obvious
fact that
the winding number   changes by $\pm 1$
every time $\Phi$ crosses zero. The sign is plus or minus,
according to whether the origin
lies to the left or right of an observer moving along the
curve $\Phi(x)$ at a time $t$ just before crossing.
 The existence
and stability of winding solitons depends
of course
on the difficulty of
crossing. This is a dynamical issue about which eqs. (2) and (3)
have nothing to say.

 We must instead turn our attention  to the field equations.
It is
  straightforward to check that
they admit the following set of  solutions:
$$
 \Phi_s(x,t) =  F_s(\omega,N)
  \  exp\Bigl({  i N x \over L} + i \omega t \Bigr) \ \ ,
\eqno(4)$$
where
$$ F_s(\omega,N) =  \sqrt{v^2+ {\omega^2\over \lambda}-
{  N^2\over \lambda L^2}}\ .
\eqno(5)$$
These describe
solitons that wind $N$ times around the mexican hat,
while  rotating with angular velocity $\omega$.
In the $\lambda\rightarrow\infty$
limit they reduce to the
  momentum and winding modes familiar
 from torroidal compactifications
of string theory.
For finite values of $\lambda$, on the other hand,
   winding and   angular momentum compete in pushing
the magnitude of the field closer or further away from the origin,
and the solutions are only defined
if $\vert N\vert < L \sqrt{\lambda v^2 + \omega^2}$.
Their energy reads
$$ {\cal E}_s = \pi  v^2 L (\omega^2
 + {N^2\over L^2}) + {\pi L \over 2\lambda}
(3 \omega^4  - {2N^2 \omega^2\over L^2} - {N^4\over L^4}) \ . \eqno(6)$$
The static winding solitons are obtained by setting
  $\omega = 0$ in the above expression. They have vanishing $U(1)$
charge, $Q = \int dx [-{i\over 2}\Phi^* \partial_t \Phi + h.c.] =
2\pi L F_s^2\omega$,
and should not in particular be confused with
 the class of non-topological solitons
known as $Q$-balls \cite{Q}.
We will concentrate mostly on them in
 the sequel, and will refer to their
energy as mass, $M_s(N) = {\cal E}_s(N,\omega=0)$.

Note in passing that had we used $\mu^{-1}$ rather than $L$
as the infrared
cutoff, we would have obtained solitons resembling those of
the sine-Gordon model. For instance
 in the $\lambda\rightarrow\infty$ limit,
the soliton with minimum winding number ($N={1\over 2}$)
is given by
$$ \Phi_s =  v \exp\Bigl( i{\pi\over 2} +
 2i tan^{-1}\exp(2\mu x)\Bigr)\ ,
 \eqno(7)$$
and has a mass $M_s = 4v^2\mu$. In
 contrast to solution (4) which we
refer to as "soliton"  by an abuse of
language common to string theorists,
solution (7) truly deserves this name,
 since it represents a non-dissipative
lump of energy localized in space.
Readers who prefer to deal with localized lumps are
  invited to
translate mentally our analysis below to a sine-Gordon-like context.

\def\baselinestretch{1.2}
\baselineskip 16 pt
\noindent
\section{
 Classical Stability and the Semiclassical Limit}

\hskip 0.6cm
 As $\lambda$ varies between $\infty$
and $({N \over vL})^2$, the static
($\omega = 0$) solutions  eqs.(4) and (5),
extrapolate continuously between
the topologically-stable winding solitons
  and the unstable vacuum, $\Phi = 0$.
  There must therefore exist some critical quartic coupling
at which they become classically
unstable, and which we will now determine.
To this end we must study
 quadratic energy fluctuations.
Let $\Phi \equiv F e^{i\Theta}$
 define polar coordinates for the complex scalar field,
 and let
$$ \eqalign{
F(x) = &F_{s}(N) + a_0 + \sum_{n=1}^\infty (a_n e^{
i n x/L} + c.c.) \cr
{d\Theta\over dx}  =& {N\over L} + {1 \over F_s(N) L}\
\sum_{n=1}^\infty (b_n e^{ i n x/L}
+ c.c.)      \cr}  \eqno(8)$$
be the Fourier decomposition of arbitrary
 space-dependent fluctuations
around the static solution.
Notice that quantization of winding forbids a continuous zero
mode in the second line.
   Plugging eq. (8)
 in  the expression for the energy,
 and keeping
up to quadratic terms in the fluctuations we
 find
$$ \eqalign{
{\cal E} - M_{s} = 2\pi L \lambda  F_{s}^2 &  a_0^2 \cr &+
{2\pi\over L} \sum_{n=1}^\infty (a_n^* \ \ b_n^*)
\left( \matrix{   n^2
 +2\lambda F_{s}^2 L^2 \  & {2N}
\cr
 {2N}   & 1 \cr}\right)
\Bigl( \matrix{ a_n \cr  b_n\cr}\Bigr) \ . \cr}
\eqno(9)$$
It is   straightforward to check that the first
  negative mode of this functional occurs in the  $n=1$ sector.
The condition for classical stability
thus reads: $ 1 + 2\lambda F_s^2 L^2 -4N^2 > 0$, which combined with
eq. (5) implies
$$  m^2 L^2 >  6 N^2-1  \ \ .
\eqno(10)$$
We conclude that for
any given winding number $N$,
classical stability imposes  a lower bound on the
  radial ("Higgs") mass measured in units of the soliton size.
Conversely, it puts an upper bound on the
allowed winding numbers
of stable solitons for any given value of $mL$.
Yet another rewriting of this constraint
can be  obtained by trading the parameter
 $L$ for the soliton mass and
the quartic coupling.
In the single-winding sector this leads to
 $$  m^3 >  {5\sqrt{5}\over 2\pi}\lambda M_{s}  \ .
\eqno(11)$$
Except for the precise numerical constant in front, we could
have obtained such an estimate by comparing the gain in gradient
energy to the loss in potential energy, if we tried to reduce the
magnitude of the soliton field in a substantial fraction of its size.
Note also that larger solitons
 are lighter and more easy to stabilize.

Even if classically stable, there is no a priori
 guarantee that our
  solitons   approximate closely true quantum
states.
 In order to address this issue,
we use a standard rescaling argument \cite{Col} that
elucidates the role of the two dimensionless parameters
  of the model.
We let
$x\rightarrow {x\over L}$,
 $t\rightarrow {t\over L}$ and
$\Phi\rightarrow {\Phi\over \sqrt{\lambda}L}$, so that space is
periodic with period $2\pi$, and the action
takes the form
$$ S =  {1\over \lambda L^2} \
 \int dt \int_0^{2\pi} dx \Bigl[ {1\over 2}
\partial_\mu \Phi^* \partial^\mu \Phi -
{1\over 4}
(\Phi^* \Phi - \lambda L^2 v^2)^2 \Bigr] \ .\eqno(12)$$
It is  now clear that the classically-relevant parameter
is $mL$, which explains why it alone was constrained by the
stability requirement, ineq. (10).
The parameter $\lambda L^2$ on the other hand plays
 the role of Planck's constant,
  so that
the semiclassical limit
($\hbar\rightarrow 0$)  amounts to taking
$$ \lambda L^2 \rightarrow 0 \ , \ \ \ {\rm with} \ \
m L  \ \ {\rm fixed} \ . \eqno(13)$$
We see that contrary to
the naive intuition, stable winding solitons can exist
even (though  not exclusively)
for arbitrarily-weak quartic coupling.
Furthermore in this limit we expect the
quantum solitons to resemble closely
their classical ascendants.

To simplify
 notation we will set in the sequel $L=1$, and work
in units of the soliton size.
We can express the
static soliton and its mass in the form:
$$ \Phi_s(x) = {m\over \sqrt{\lambda}}
 \sqrt{{1\over 2} - {N^2\over m^2}}
\exp (iNx) \ , \eqno(14) $$
and
$$ M_s = {\pi N^2 m^2 \over 2\lambda}
(1 - {N^2\over m^2}) \ , \eqno(15)$$
to
exhibit the large  mass and
 amplitude, characteristic of classical solutions in
theories with weak coupling.
In the semiclassical approximation the
  zero mode, corresponding to global phase  rotations:
$\Phi_s \rightarrow \Phi_s e^{i\theta}$,
can be
quantized separately from the remaining degrees of freedom.
Its
  conjugate momentum is the $U(1)$ charge $Q$,
  which is forced to take integer values.
One can easily check that a fixed charge
does not affect the
  amplitude and mass of the soliton to lowest order in the
semiclassical expansion.

\def\baselinestretch{1.2}
\baselineskip 16 pt
\noindent
\section{  Quantum decay}

\hskip 0.6cm  Assuming they are classically
 stable, winding solitons will eventually decay by
 quantum tunneling,  much like a ribbon would slip off the
tip of a hat under the influence of external noise.
The decay rate can be obtained from the imaginary part of the
amplitude for the soliton ground state
 to evolve to itself in Euclidean
time. Assuming a small imaginary part we can
 express the decay width as follows:
 $$\Gamma \simeq  \lim_{T\rightarrow\infty}
 -{1\over T} e^{M_s T} Im <s\vert
e^{-HT}\vert s > \ . \eqno(16)$$
 Since in its ground state
the soliton has zero angular velocity, its wavefunction
  does not
depend on the overall orientation $\theta$,
so that
  $\vert s> = \int {d\theta\over 2\pi} \vert \theta>$.
To evaluate the amplitude in eq. (16) we must therefore
sum over all Euclidean histories which take
the soliton configuration   back to itself, up to an overall
global phase. In the semiclassical approximation the relevant
contributions will come from instantons, with the above
boundary conditions, and with an {\it odd} number
of negative modes in the quadratic fluctuations around them
\cite{CC}\cite{Col}.
 The latter requirement ensures that the determinant
of these fluctuations is negative, so that the instanton contributes
to the imaginary part of the amplitude.
The expression for the
decay width takes the form
$$ \Gamma \simeq \lim_{T\to\infty}
 \Gamma_0 exp{ [- \overline{S}  + M_s T] }\ ,
\eqno(17) $$
 where $\overline{S}= {1\over \lambda}
f(m) $ is the   Euclidean action of the
  instanton, while the prefactor
$\Gamma_0$ is a characteristic mass scale of the model
 arising from the fixing of the time-translational zero-mode, the
ratio of (constrained) determinants
around the instanton and the static soliton, and
 the wavefunction of the soliton at
a given value of $\theta$ \cite{Col}.
We will be interested in exponential accuracy,
 and wont worry about the
prefactor $\Gamma_0$.

 We have found one instanton which contributes to the
above decay rate. It is given by
$$ \overline{\Phi}(x,\tau) \equiv {\overline F}
e^{i{\overline\Theta}}  = \Phi_{s}(x) \tanh
\Bigl[ A
 (\tau-\tau_0)  \Bigr] \ \ ,
\eqno(18a)$$
where
$$A = \sqrt{\lambda\over 2} F_s =
{m\over 2} \sqrt{1-{2N^2\over m^2}} \ .
\eqno(18b)$$
This instanton
 describes the motion of a ribbon, slipping towards the tip of
the mexican hat  before
slipping back to its original
 position rotated through an angle $\pi$.
It  is centered at $\tau_0$, and has a size $\sim m^{-1}$.
The difference of its action from that of a static soliton is finite
and  equal to
$$ \lim_{T\to\infty} [\overline{S}-M_s T] =
 {\pi m^3\over 2\lambda} \Bigl(1 - {2 N^2\over m^2}\Bigr)^{3\over 2}
\ . \eqno(19)$$
To prove that the number of negative modes is odd, we consider
   quadratic
 fluctuations about this solution. We decompose
a generic Euclidean history into space-Fourier modes as follows:
$$ \Phi(x,\tau) = {\overline\Phi} + \Biggl[
a_0(\tau) + \sum_{n=1}^{\infty} \Bigl( a_n(\tau) exp({inx}) +
b_n^*(\tau) exp({-i nx}) \Bigr)
 \Biggr] exp({iN x })\ , \eqno(20)$$
and concentrate for convenience  on the $N=1$ single-winding soliton.
Plugging the decomposition (20)
into the Euclidean action,  and keeping only
the quadratic terms in the fluctuations we find:
$$\eqalign{ S^{(2)} = 2\pi  \int d\tau \
 \Biggl[
(a_0^R \ \ a_0^I)&
\left( \matrix{   L_0 + {\lambda \overline{F}^2/ 2}
   \  & 0
\cr
 0   &  L_0 - {\lambda \overline{F}^2/ 2}   \cr}\right)
\Bigl( \matrix{ a_0^R \cr a_0^I\cr}\Bigr)
\cr \ & \ \cr
 &+
\sum_{n=1}^\infty
(a_n^* \ \ b_n^*)
\left( \matrix{  L_n \  &   {\lambda \overline{F}^2/ 2}
\cr
 {\lambda \overline{F}^2/ 2}   & L_{-n} \cr}\right)
\Bigl( \matrix{ a_n \cr  b_n\cr}\Bigr)
 \Biggr] \cr}
\ , \eqno(21a)$$
where $a_0^R$ and $a_0^I$ are the real and
 imaginary components of $a_0$, and
the differential operator $L_n$ is given by
$$ L_n = -{1\over 2} {d^2\over d\tau^2} +
 {(n+1)^2\over 2 } + {\lambda\over 2}
(2\overline{F}^2 - v^2) \ . \eqno(21b)$$
Notice that
 for $n>0$, all eigenvectors come in degenerate complex conjugate
pairs, so the number of negative eigenvalues from these sectors is even.
We concentrate therefore on the operator in the $n=0$ sector,
which
can be expressed as follows:
$${ 1\over 2}
\left( \matrix{  -{d^2/ d\tau^2} -
{6 A^2/\cosh A\tau} + 4A^2  \
&   0 \cr
 0  &   - {d^2/ d\tau^2} -
2{ A^2/ cosh^2 A\tau} \cr} \right)
\ . \eqno(22)$$
This operator has a well-known discrete
spectrum
\cite{Landau},   consisting of one negative eigenmode
$\psi_-= ( 0 \ , {1/ \cosh A\tau})$ with
 eigenvalue $-{A^2/ 2}$, and
one zero eigenmode $\psi_0= ({1/ cosh^2 A\tau} ,
 \ 0)$ ,   corresponding to
time translations of the instanton
 \footnote{There is another zero eigenmode
in the $n=0$ sector
given by $(0\ , \tanh A\tau)$. It corresponds to
global phase rotations of the instanton,
but  is clearly not normalizable.}.
All other eigenvalues are positive,
and this concludes the proof that the instanton
eqs. (18a,b) contributes to the decay width of the soliton.

A straightforward variational argument shows in fact
that   the number of
 negative modes is greater than one and  growing  as
  $m\rightarrow\infty$.
 This could be indicating that our instanton
  is not the dominant saddle point.
Intuitively we may expect that   pulling the ribbon over the tip
of the hat on one side only, is a preferable decay mode
  for solitons of
large size. In any case the estimate, eqs. (17) and (19), provides
at worse an upper bound on the soliton's lifetime.

 \def\baselinestretch{1.2}
\baselineskip 16 pt
\noindent
\section{  Gauging and a Second Higgs}

\hskip 0.6cm
Let us consider next what happens when the complex scalar field(s)
are coupled to a $U(1)$ gauge boson.
This so-called Abelian-Higgs model has been
 employed frequently  in the past
  in order to study the
non-perturbative structure of gauge theories. The action    is
$${\cal S} = \int dt
 dx \Bigl[ -{1\over 4} F_{\mu\nu}F^{\mu\nu}
 + \sum_J
{1\over 2}  \vert (\partial_{\mu} -
ie_J A_\mu) \Phi_J \vert^2
 - V(\Phi_J)  \  \Bigr] \ \ ,\eqno(23)$$
with $e_J$  the $U(1)$ charge of $\Phi_J$.
We restrict first ourselves to a	single scalar
with the mexican-hat potential of eq. (1). To look for static solitons
we go to the $A_t = 0$ gauge and consider time-independent configurations.
Since for these
  $F_{tx}=0$, the field
    equations reduce  to
$$\Phi^* (\partial_x - ieA_x) \Phi =
{dV\over d\Phi} = 0 \ . \eqno(24)$$
These have no solutions
other than the vacuum ($\Phi = v$, $A_\mu = 0$)
 and its gauge transforms.
The winding solitons of the ungauged model have all
disappeared, even in the $\lambda\to\infty$ limit.

The reason for this novel instability is easy to explain.
  Although  the winding number of the scalar
field may  still be conserved by the time evolution, there
now exist winding {\it vacua}
( $\Phi = v e^{iNx} \
,\  A_x = {N\over e}$) for any given value of $N$.
By turning
on the $A_x$ gauge field, we can
  thus deform continuously any winding configuration
to the corresponding vacuum  plus oscillations.
Furthermore,
because a static $A_x$ gauge field costs zero energy
in one space dimension,
 there is no energy barrier hindering such a process
\footnote{
This conclusion
gets modified for higher $d$,
for which  classically stable solitons   exist
  below some critical gauge coupling \cite{BT}.}.
Another way
 of saying this   is by noting that
the current, eq. (2), is topologically-conserved but gauge-variant.
The corresponding winding charge changes
under large gauge transformations,
which allow us to bring any configuration back to the
zero-winding sector.
A gauge-invariant current does exist:
$${\tilde J}^\mu = {1\over 2\pi i}
\epsilon^{\mu\nu}( \partial_\nu
{\sl log}\Phi - ieA_\nu) \ , \eqno(25)$$
 but it has
an anomaly,
 $$\partial_\mu {\tilde J}^\mu = - {e\over 4\pi} \epsilon_{\mu\nu}
F^{\mu\nu} \ , \eqno(26)$$
even in the $\lambda\rightarrow\infty$ limit.
The corresponding charge is not conserved by the gauge-field
dynamics.

This situation changes drastically if the theory has more than one
  charged scalars.
Consider for instance the simplest case of two
fields, both with frozen
magnitudes  and with equal $U(1)$ charges.
Scalar configurations
 are now labelled by two winding numbers, $N_1$
and $N_2$, but winding vacua only exist in the $N_1=N_2$
sectors.
Sectors with non-zero relative winding ($N\equiv N_1-N_2\not= 0$)
cannot be transformed to the trivial sector and
are in this case  guaranteed to contain stable solitons.
It is easy, in fact,  to check that $N$ is the charge of
a gauge-invariant and anomaly-free current, since it is
the winding number of the gauge-invariant composite field
$\Phi_1^*\Phi_2$.

Let us relax now the freezing of
 the magnitudes of the two scalar fields
and
repeat the analysis of the previous sections.
 To keep our expressions simple, we
restrict  ourselves to  the generic
    $U(1)\times U(1)\times Z_2$-symmetric potential
\footnote{We can again
 trade  the infrared cutoff $L$ for a gauge-invariant mass
term $\delta V = \mu^2  Re[\Phi_1^* \Phi_2]$,
 which lifts the accidental $U(1)$ symmetry of the potential.
This symmetry on the other hand makes the technical analysis much
more pleasant.}
,
 $$\eqalign{
 V(\Phi_1, \Phi_2) = {\lambda\over 4}&
(\vert \Phi_1\vert^2 - v^2)^2
 + {\lambda\over 4} (\vert\Phi_2\vert^2 - v^2)^2
 + {{\lambda^\prime}\over 2}
(\vert \Phi_1\vert^2 - v^2)( \vert \Phi_2\vert^2 - v^2)\cr
&= {\lambda + \lambda^\prime\over 8} ( \vert \Phi_1\vert^2 +
\vert \Phi_2\vert^2 - 2 v^2)^2
+ {\lambda - \lambda^\prime\over 8}
 ( \vert \Phi_1\vert^2 - \vert \Phi_2\vert^2
)^2 \cr}
 . \eqno(27)$$
Assuming $\lambda > \vert \lambda^\prime \vert $,
 we can choose the vacuum
    at $\Phi_1 = \Phi_2 = v$. The perturbative
spectrum consists of two
massless goldstone bosons (one of which is eaten by the longitudinal
component of the photon), and two massive scalars with
$$ m_{\pm}^2 = 2v^2 (\lambda\pm \lambda^\prime) .
 \eqno(28)$$
The time-independent field equations admit the following set of solutions:
$$  \Phi_{J,s}(x) = \sqrt{ v^2-{N^2\over 4(\lambda +
\lambda^\prime) }} \  exp\Bigl({  i N_J x }\Bigr) \ \
\ \ (J=1,2)  \ , \eqno(29a)$$
$$ A_x =  {N_1 + N_2\over 2e}\ \ ,
\eqno(29b)$$
  subject to the constraint
$$ m_+^2 > {N^2\over 2} \ .\eqno(30)$$
The mass of these   solitons reads
$$ M_{s} = {\pi \over 2  } N^2
 \Bigl[v^2 - { N^2\over 8(\lambda
 +\lambda^\prime) }\Bigr]\ \ ,\eqno(31)$$
and is only a function of the relative winding $N$ as expected.
Indeed, solutions
   with the same value of  $N$ can be transformed to each other
  by large gauge transformations. Strictly-speaking at the full
quantum level the solitons are linear superpositions of
all these gauge-transforms,
i.e. excitations above the appropriate $\theta$-vacuum of the theory.

 In order to check classical stability
let us
Fourier decompose the fluctuations of
appropriate  gauge-invariant combinations of fields, around the
solution (29),
 as follows:
$$ \eqalign{
F_J   = F_s + & a^{\ J}_0 +
\sum_{n=1}^\infty (a^{\ J}_n e^{
i n x} + c.c.) \cr
{d\Theta_J\over dx}- e A_x
 &= \pm { N\over 2 } + {1\over F_s} \Bigl[ b_0 +  \
\sum_{n=1}^\infty (b^{\ J}_n e^{ i n x}
+ c.c.) \Bigr] \ .    \cr}  \eqno(32)$$
Here $F_s$
is the common magnitude of the two scalar fields $\Phi_{J,s}$,
  the sign in the second line is
 plus or minus for $J=1$ and $2$ respectively,
while the constant mode $b_0$
comes entirely from the gauge field, and is therefore the same for
both values of $J$. Plugging (32) into the expression for
the energy, and keeping  up to quadratic terms in the fluctuations
we find
$$ {\cal E} - M_{s} =
{2\pi } \sum_{n=0}^\infty v_n^{\dag}
{\cal E}^{\prime\prime}_n v_n \ \ , \eqno(33)$$
where
$$  v_0 \equiv \left(\matrix{ a_0^{\ 1}
 \cr a_0^{\ 2} \cr b_0 \cr} \right) \ \ ;
 \ \ {\cal E}^{\prime\prime}_0  \equiv  {1\over 2}\
\left( \matrix{
 2\lambda F_s^2   &
 2\lambda^\prime F_s^2   &
 N
\cr
  2\lambda^\prime F_s^2  &
 2\lambda  F_s^2   &
 -N
\cr
N  &
 -N &
  2\cr}\right)
 \ , \eqno(34)$$
and
$$ v_n \equiv \left( \matrix{ a_n^{\ 1}
\cr a_n^{\ 2} \cr b_n^{\ 1} \cr b_n^{\
2} \cr} \right) \ \  ; \ \
{\cal E}^{\prime\prime}_n \equiv
\left(\matrix{
n^2 + 2\lambda F_s^2   &
 2\lambda^\prime F_s^2  &
 {N    }& 0
\cr
 2\lambda^\prime F_s^2   &
n^2 + 2\lambda F_s^2  & 0 &
-{N  }
\cr
{N  }& 0 &  1 & 0 \cr
0 & -{N  }&  0 & 1  \cr}\right)
  \eqno(35) $$
for $n>0$.
The three eigenvalues of the matrix  ${\cal E}^{\prime\prime}_0$ read
$$ \eqalign{ X_1 & = (\lambda + \lambda^\prime) F_s^2 \cr
X_{2,3} & = {1\over 2} \Bigl[ 1 + (\lambda-\lambda^\prime)F_s^2 \Bigr]
\pm {1\over 2} \sqrt{\Bigl( 1 -
 (\lambda-\lambda^\prime)F_s^2\Bigr)^2 + 2N^2}
\ .   \cr}
\eqno(36)$$
Only $X_3$ is potentially negative, leading to the stability
condition $ 2F_s^2 (\lambda - \lambda^\prime) > {N^2   }
 $. Turning now to the $n>0$ sectors,
we note first that it is sufficient to impose that
 ${\cal E}^{\prime\prime}_1$  be a positive-definite matrix.
 Its eigenvalues are
the solutions to the two quadratic equations
$$ X^2 - 2[  1 +   F_s^2
(\lambda\pm \lambda^\prime)] X +
[1 - N^2 + 2  F_s^2
(\lambda\pm \lambda^\prime)] = 0 \eqno(37)$$
They are all positive   provided
$ 2F_s^2 (\lambda\pm \lambda^\prime) > {N^2-1  }
 $.
 Using our expression for $F_s$, we can finally summarize the
  conditions for classical stability as follows:
 $$ m_{-}^2 \Biggl( 1-{N^2\over 2m_+^2}\Biggr)
  > N^2 \ \ , {\rm and} \ \
 m_{+}^2 \Biggl( 1-{N^2\over 2m_+^2}\Biggr)  > N^2-1 \ .\eqno(38)$$
We must of course also assume ineq. (30), which was the
condition for the existence of the classical solutions.

A surprising feature of these constraints is that they can be
satisfied in the sector $N=1$, by
 winding solitons lying arbitrarily close to
the tip of the mexican-hat potential! More generally,
contrary to what happened in the single-scalar model,
  gauging seems to have
a stabilizing effect when the Higgs sector is extended.
This can be  also seen from the expression for the
energy: for fixed $N$, $v$ and $\lambda\to\infty$,
the mass in the ungauged model is double the mass given by eq. (31).
  Roughly-speaking,   the solitons of the gauged model,
 can be thought off as a bound pair of one soliton and one antisoliton,
which split in two the total winding number thus reducing the total energy.

\def\baselinestretch{1.2}
\baselineskip 16 pt
\noindent
\section{  Conclusions }

 \hskip 0.6cm
We have analyzed a new class of (quasi-topological)
solitons in theories
 with a topologi- cally trivial target space.
They owe their stability
to the dynamical exclusion of some region of the target space,
and become topologically-stable in the limit
of infinite coupling. They
  continue, however,  to exist even
  in the semiclassical limit of weak coupling, where furthermore their
properties can be reliably calculated.
Such winding solitons
  decay   by quantum tunneling, so they have exponentially-long
but finite lifetimes.
One intriguing consequence of this fact is that Skyrmions could decay
in the linear Gell-Mann-Levy model \cite{BT}.
Turning on (sufficiently strong) gauge
 interactions destabilizes these solitons,
unless the scalar sector is extended. The application of these ideas
to the Higgs sector of the standard electroweak model \cite{BT} was
in fact one of the main motivations for the present work.
Another possible
application  is in the context of string theory.
One can think of the  mexican-hat potential of eq. (1)   as a
tachyon background which effectively traps low-lying fundamental strings
in one compactified dimension.
The possibility of analogous
 effects in gravitational
backgrounds, as well as the role of the
 breaking of world-sheet scale invariance
in this context, deserve  further study.

\vskip 0.3cm
{\bf Aknowledgements}
C.B. thanks the Research Center of Crete
and the Physics Deptartment of the
University of Crete, and T.N.T.
thanks the Centre de Physique
Th\'eorique of the Ecole Polytechnique
for hospitality while this work was being
completed. This research was supported in part by EEC grants
SC1-0430-C and   CHRX-CT93-0340.


\newpage



\begin{thebibliography}{6666}



\bibitem{Skyrme} T.H.R. Skyrme,
 Proc. Roy. Soc. (London), Ser. {\bf A260}
(1961) 127.

\bibitem{bubbles}
 A.P.Malozemoff and J.C.Slonczewski,
{\it Magnetic Domain Walls in Bubble
Materials}, Academic Press (1979); \\
  N.Papanicolaou and T.N.Tomaras,  Nucl.Phys.
 {\bf B360} (1991) 425,  and references
therein.


\bibitem{CS}
E. Cremmer and J. Scherk, Nucl. Phys. {\bf B103} (1976) 399.

\bibitem{Q} T.D. Lee, {\it  Particle Physics and Introduction to Field
Theory},  Harwood Academic Publishers (1981).

\bibitem{BT}  C. Bachas and T.N. Tomaras, Higgs-sector solitons,
Ecole Polytechnique and
Crete-preprint to appear, and references therein.


\bibitem{Col} S. Coleman,
{\it Aspects of Symmetry}, Cambridge U. Press (1985).


\bibitem{CC} J.S. Langer,
 Ann. Phys. (New York) {\bf 41} (1967) 108;\hfil\break
S. Coleman, Phys. Rev. {\bf D15} (1977) 2929; \hfil\break
C. Callan and S. Coleman, Phys. Rev. {\bf D16} (1977) 1762 .


\bibitem{Landau} L.D. Landau and E.M. Lifshitz,
 {\it Quantum Mechanics}, Pergamon
Press.

\end{thebibliography}
\end{document}